\documentclass[3p,times]{elsarticle}

\usepackage{ecrc}
\usepackage[final]{epsfig}

\volume{00}

\firstpage{1}

\journalname{Nuclear Physics A}

\runauth{}

\jid{nupha}

\usepackage{amssymb}
\usepackage[figuresright]{rotating}

\usepackage[belowskip=-15pt,aboveskip=0pt]{caption}

\begin{document}

\begin{frontmatter}



\dochead{}

\title{Non-perturbative Heavy-Flavor Transport at RHIC and LHC}


\author{Min~He\fnref{label0}}
 \fntext[label0]{Speaker,\,\, {\it Email address}: mhe@comp.tamu.edu }
\author{Rainer~J.~Fries, and Ralf~Rapp}
\address{Cyclotron Institute and Department of Physics and
Astronomy, Texas A\&M University, College Station, Texas 77843-3366, U.S.A.}

\begin{abstract}
We calculate open heavy-flavor (HF) transport in relativistic heavy-ion collisions by
applying a strong-coupling treatment in both macro- and microscopic dynamics (hydrodynamics and
non-perturbative diffusion interactions). The hydrodynamic medium evolution is quantitatively
constrained by bulk and multi-strange hadron spectra and elliptic flow. The heavy quark (HQ)
transport coefficient is evaluated from a non-perturbative $T$-matrix approach in the
Quark-Gluon Plasma (QGP) which, close to the critical temperature, leads to resonance
formation and feeds into the recombination of heavy quarks on a hydrodynamic hypersurface.
In the hadronic phase, the diffusion of HF mesons is obtained from effective hadronic
theory. We compute observables at RHIC and LHC for non-photonic electrons and HF mesons,
respectively.
\end{abstract}


\end{frontmatter}

% \linenumbers

\section{Introduction}
\label{intro}
%%%%%%%%%%%%%%%%%%%%%%%%
A deconfined state of nuclear matter has been predicted by large-scale numerical simulations
of Quantum Chromodynamics (QCD) on the lattice~\cite{Aoki:2006we,Bazavov:2011nk}.
Utilizing ultra-relativistic heavy-ion collision (URHIC) experiments, this new phase
%, which presumably prevailed in the very early Universe,
has been identified as a strongly coupled Quark-Gluon Plasma
(sQGP)~\cite{Adams:2005dq,Adcox:2004mh}. With the advent of LHC, the endeavor of
characterizing the properties of sQGP has entered a new stage~\cite{Aamodt:2010pb,Aamodt:2010pa}.

Heavy quarks, due to their large masses ($m_c\simeq1.3$\,GeV for charm
and $m_b\simeq 4.2$\,GeV for bottom), are produced in primordial hard
collisions~\cite{Adler:2004ta}, and their number is expected to be conserved in the
subsequent evolution of the medium. Through interactions with medium constituents,
the spectra of HQs are modified, yet they may not fully thermalize. This
makes them valuable probes of the hot and dense matter created in URHICs~\cite{Rapp:2009my}.
The suppression of non-photonic electrons ($e^\pm$) from HF decays measured
in Au+Au($\sqrt{s_{\rm NN}}=200$\,GeV) at RHIC~\cite{Abelev:2006db,Adare:2006nq,Adare:2010de} is
surprisingly large, and their elliptic flow~\cite{Adare:2006nq,Adare:2010de}
indicates a substantial collectivity of charm quarks. It is difficult to explain these
results within perturbative QCD
approaches~\cite{Armesto:2005mz,Gossiaux:2008jv,Alberico:2011zy,Uphoff:2011ad} for HQ
scattering in the QGP. They rather indicate the need for a non-pertubative treatment
of HQ diffusion in the low- and intermediate-momentum region~\cite{vanHees:2005wb}.

Here, we report on applications of our recently developed non-perturbative
framework for open HF transport in medium~\cite{He:2011qa} to $D$- and $B$-meson
observables at LHC, and predict upcoming RHIC data for
flavor-separated $e^\pm$ spectra.

%%%%%%%%%%%%%%%%%%%%%%%%%%%%%%%%%%%%%%%%%%%%%%%%%%%%%%%%%%%%%%%
\section{Theoretical Framework: HQ Transport and Hadronization}
\label{theo}
%%%%%%%%%%%%%%%%%%%%%%%%%%%%%%%%%%%%%%%%%%%%%%%%%%%%%%%%%%%%%%%
%\subsection{Transport coefficient}
The kinetics of HQs in the QGP can be described by Fokker-Planck dynamics,
which in practice is simulated through relativistic Langevin diffusion~\cite{Rapp:2009my}.
The required transport coefficient is the HQ thermal relaxation rate, $A(p,T)$.
We employ the results from a thermodynamic $T$-matrix approach~\cite{Riek:2010fk} for
HQ-light quark scattering with input potentials taken as the HQ internal energy computed
in thermal lattice QCD. While HQ-gluon scattering was treated perturbatively
in Refs.~\cite{He:2011qa,He:2012df}, we here adopt a recent update also computed
in the $T$-matrix formalism~\cite{Huggins:2012dj}. The $T$-matrices exhibit resonant
states close to threshold at temperatures $T\simeq 1-1.5$\,$T_c$, which accelerates the
HQ thermal relaxation by a factor of $\sim$3-5 for the HQ-light quark scattering
(factor $\sim$2 for HQ-gluon) over leading order pQCD results~\cite{Riek:2010fk}.
After hadronization, the open HF mesons continue to diffuse in the hadronic fluid.
The thermal relaxation rate of $D$ mesons has been calculated by using elastic scattering
amplitudes from effective hadronic theory~\cite{He:2011yi}. As summarized in
Ref.~\cite{He:2012df}, the charm diffusion coefficient $\mathfrak{D_s}=T/[m_{c,D}A(p=0,T)]$
is essentially continuous across the phase transition region, possibly indicative of
``quark-hadron duality". The $B$-meson thermal relaxation rate used is obtained from
that of $D$ mesons by scaling the relevant mass, assuming they have the same spatial
diffusion coefficient (in the large mass limit).

%\subsection{Diffusion and hadronization}
The Langevin simulations of HQ diffusion in QGP are implemented into the space-time evolution
of a background medium modeled by boost-invariant ideal hydrodynamics~\cite{He:2011zx}.
The equation of state (EoS) was constructed from a parameterization of recent lattice-QCD
data in the high temperature phase~\cite{Borsanyi:2010cj,Bazavov:2011nk}, combined with a
partial chemical-equilibrium EoS for the hadronic phase. A compact initial-density profile
with pre-equilibrium flow was introduced resulting in fair fits to multistrange-hadron
observables close to the transition temperature, $T_c=170$\,MeV, and to bulk-hadron
observables at $T_{\rm kin}=110$\,MeV. Further details of
the implementation of the Langevin simulation can be found in Ref.~\cite{He:2011qa}.

At $T_c=170$\,MeV, the heavy quarks are hadronized into HF mesons. We accomplish this by first
applying resonance recombination~\cite{Ravagli:2007xx,He:2011qa} with thermal light quarks
on the pertinent hydro-hypersurface. Remaining heavy quarks are treated with $\delta$-function
fragmentation (as used in the fits to baseline HQ spectra in p+p collisions). For a reliable
recombination dynamics at low and intermediate transverse momenta, $p_T$, it is crucial that
the correct thermal equilibrium limit is respected. We have verified this in the presence
of the full space-momentum correlations generated from the HQ Langevin dynamics and
hydrodynamic flow of the thermal light quarks. The relative partition between recombination and
fragmentation is determined by a coalescence probability which is obtained from the HQ
$T$-matrix scattering rate, and thus
consistent with the underlying transport coefficient~\cite{He:2011qa}. At $T_c$, the in-medium
quark and meson masses are used: $m_q=0.3$\,GeV, $m_c=1.7$\,GeV, $m_b=5.2$\,GeV, $m_D=2.1$\,GeV
and $m_B=5.6$\,GeV, with $D$- and $B$-meson widths of $\Gamma_{\rm D,B}=0.1$\,GeV.
After hadronization, the Langevin simulation resumes for $D$ and $B$ mesons in the hadronic
phase until hydrodynamic kinetic freezeout at $T_{\rm kin}=110$\,MeV.
%We have checked this procedure gives a relative strength in $D$-meson and $B$-meson
%relaxation rate that is consistent with the comparison between $c$- and $b$-quark
%relaxation rate calculated from $T$-matrix approach.

%%%%%%%%%%%%%%%%%%%%%%%%%%%%%%%%%%%%%%%%
\section{Open Heavy-Flavor Observables}
\label{obser}
%%%%%%%%%%%%%%%%%%%%%%%%%%%%%%%%%%%%%%%%
%%%%%%%%%%%%%%%%%%%%%%%%%%%%%%%%%%%%%%%%%%%%%%%%%%%%%%%%%%%%
\subsection{Flavor-separated non-photonic electrons at RHIC}
%%%%%%%%%%%%%%%%%%%%%%%%%%%%%%%%%%%%%%%%%%%%%%%%%%%%%%%%%%%%
\begin{figure}[!t]
\begin{center}
%\rule{1cm}{2cm}
\includegraphics[width=8cm]{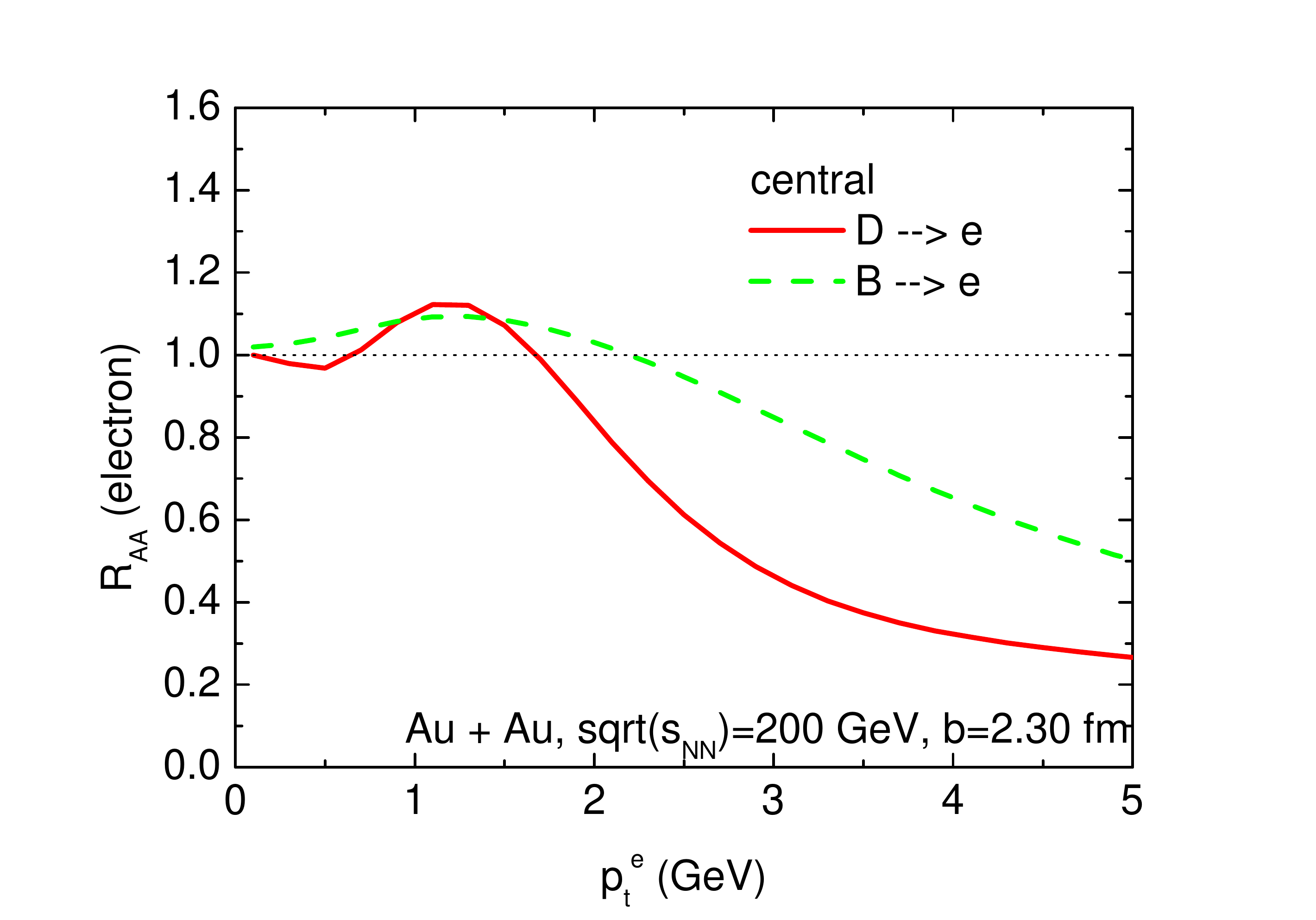}
\includegraphics[width=8cm]{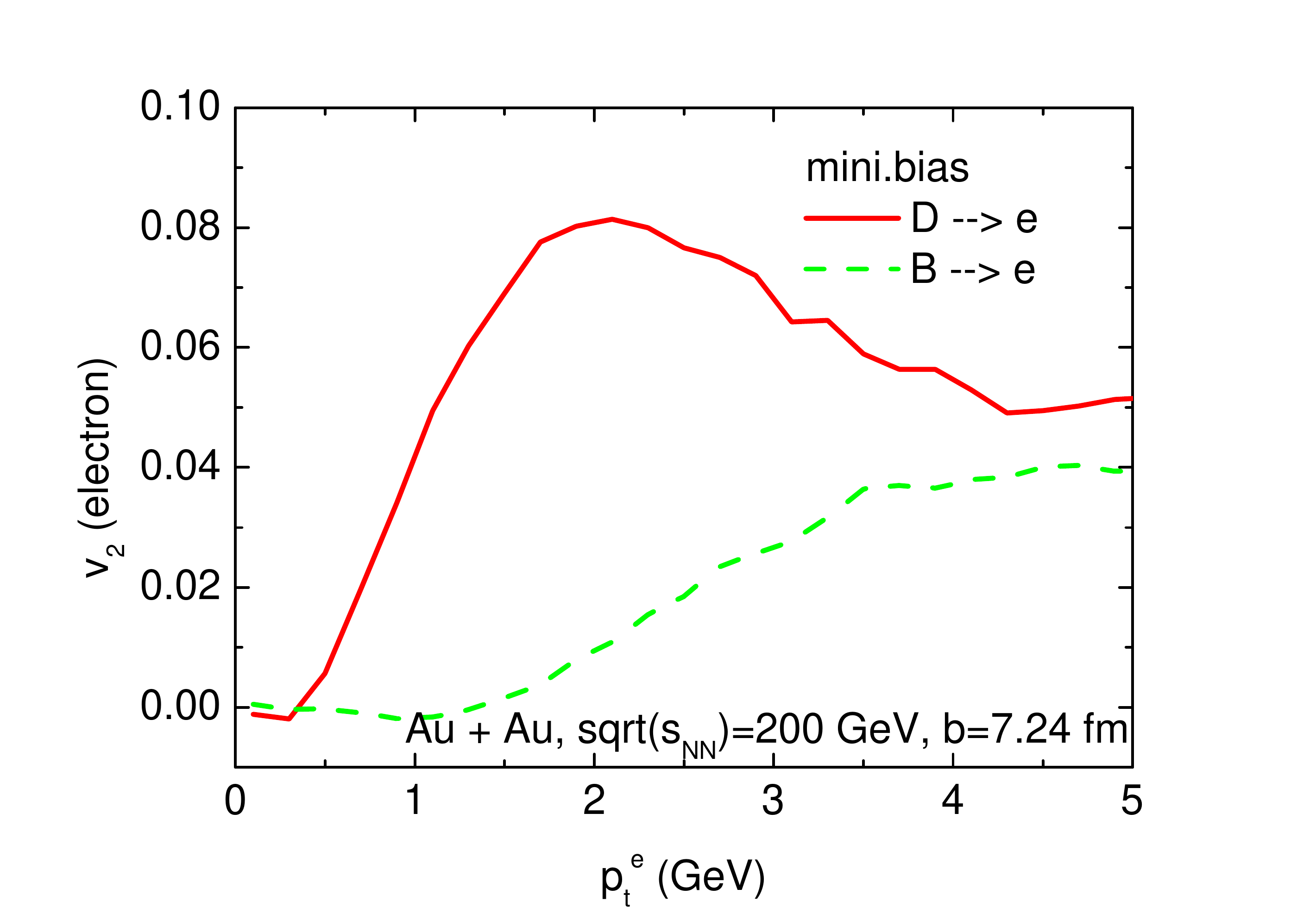}
\end{center}
\vspace{-0.2cm}
\caption{(Color online) Separate charm- and bottom-electron $R_{\rm AA}$ (left panel)
and $v_2$ (right panel) in Au+Au ($\sqrt{s_{\rm NN}}$=200\,GeV).}
\label{RHIC_electrons}
\end{figure}
The measured $e^\pm$ nuclear suppression factor, $R_{\rm AA}$, and elliptic flow, $v_2$,
for Au+Au ($\sqrt{s_{\rm NN}}$=200\,GeV) collisions thus far are superpositions of open-charm
and -bottom contributions~\cite{Abelev:2006db,Adare:2006nq,Adare:2010de}. In anticipation of
experimental results for charm- and bottom-separated $e^\pm$~\cite{Nouicer}, we present
in Fig.~\ref{RHIC_electrons} our pertinent predictions. The charm-$e^\pm$ $R_{\rm AA}$
exhibits a flow bump at $p_t^e\simeq 1.5$\,GeV, which is a combined consequence of charm-quark
thermalization in the QGP diffusion processes and the subsequent coalescence with thermal
light quarks at hadronization~\cite{He:2011qa}. Note that, from a microscopic point of view,
both processes are of identical origin (heavy-light quark $T$-matrix). At $p_T^e\simeq 5$\,GeV,
the charm $e^\pm$ suppression reaches 0.2-0.3. In accord with approaching thermalization in
the suppression factor, the charm-$e^\pm$ also acquires substantial elliptic flow, reaching
$\sim$8\% at $p_t^e\simeq 2$\,GeV. Bottom electrons suffer less suppression, at least below
$p_t^e\simeq 5$\,GeV, due to the larger $b$-quark mass. We also observe the
bottom-$e^\pm$ $v_2$ to saturate at a significantly larger $p_t^e$ than for charm, as
a result of the larger $m_b/m_q$ ratio that is operative in picking up light-quark $v_2$
through coalescence.

%%%%%%%%%%%%%%%%%%%%%%%%%%%%%%%%%%%%%%%
\subsection{$D$ and $B$ mesons at LHC}
%%%%%%%%%%%%%%%%%%%%%%%%%%%%%%%%%%%%%%%
\begin{figure}[!t]
\begin{center}
\includegraphics[width=8cm]{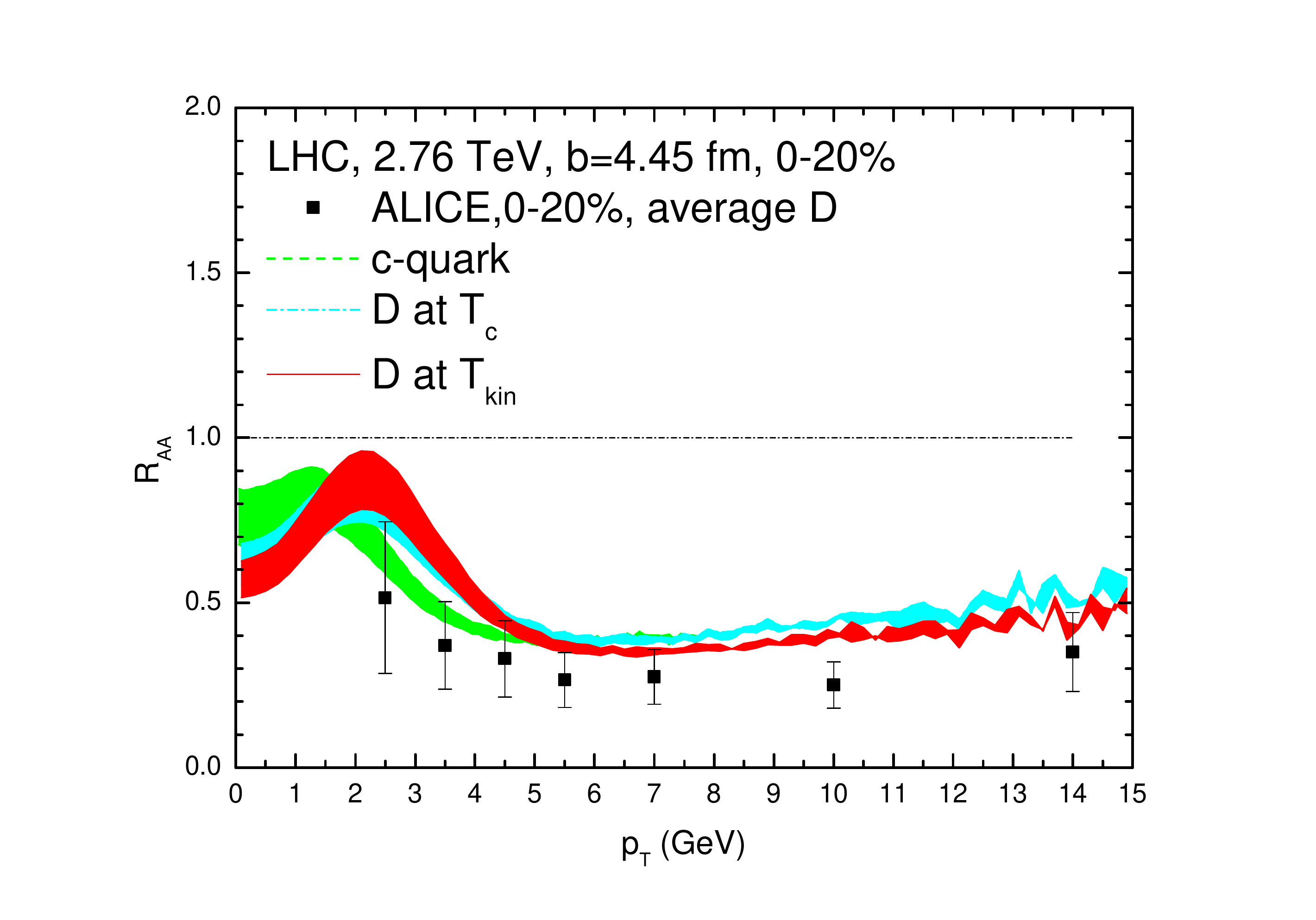}
\includegraphics[width=8cm]{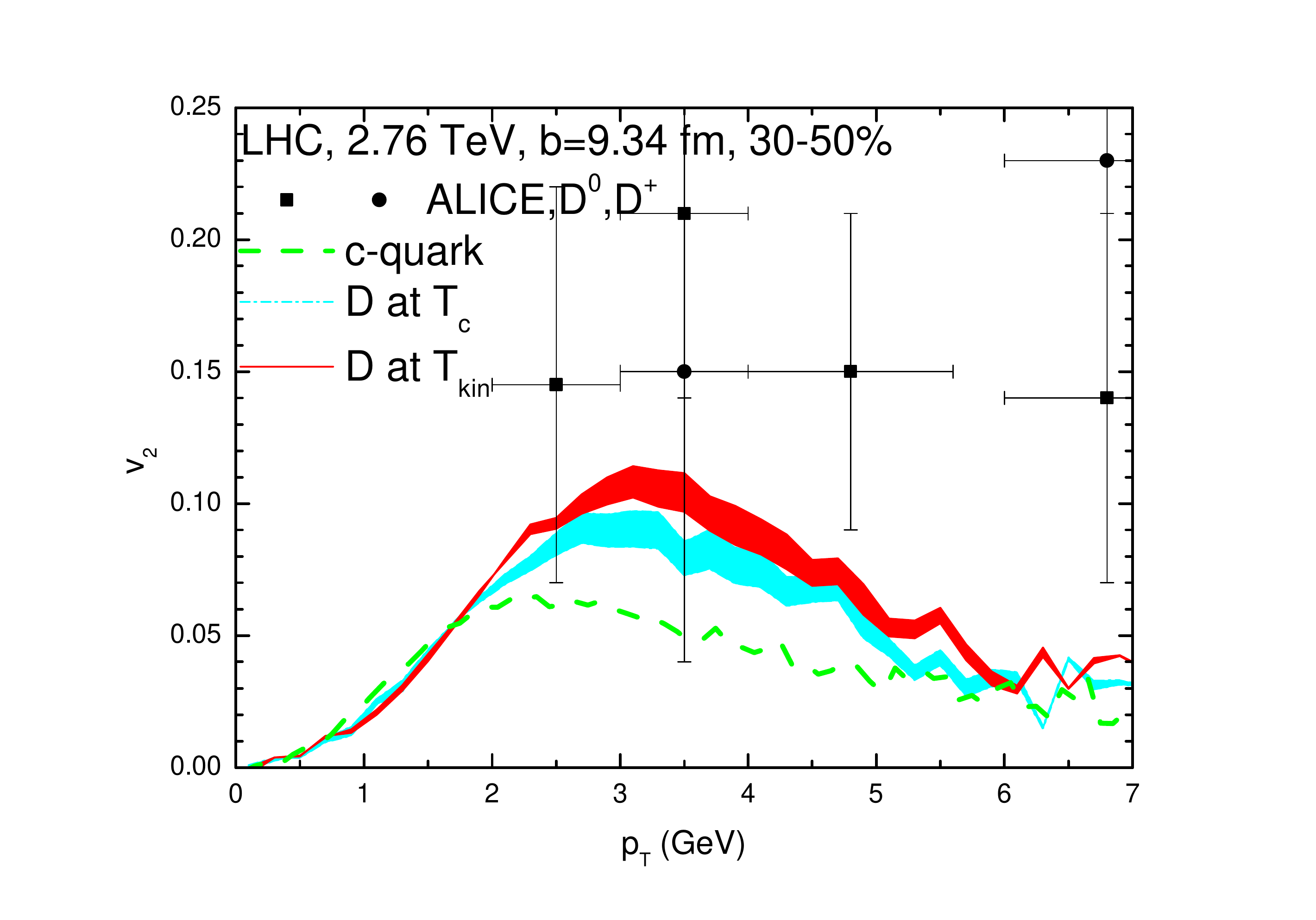}
\end{center}
\vspace{-0.3cm}
\caption{(Color online) $D$-meson $R_{\rm AA}$ (left panel) and $v_2$ (right panel) in
Pb+Pb ($\sqrt{s_{\rm NN}}$=2.76\,TeV) collisions, compared to
ALICE data~\cite{ALICE:2012ab,Giacomo_Ortona}. Our results for charm quarks, $D$ mesons
after hadronization at $T_c$ and after hadronic diffusion at $T_{\rm kin}$, are shown
separately. The bands indicate the prevalent uncertainties in each case: shadowing
(64.0\%-77.6\%) for $R_{\rm AA}$ and charm-quark coalescence probability (50\%-90\%) for
$v_2$.}
\label{LHC_D}
\end{figure}
Next, we apply our non-perturbative approach to the LHC. Toward this end, we employ FONLL
pQCD~\cite{Cacciari:1998it} $D_0$-meson and $B$-meson spectra in $\sqrt{s}=2.76$\,TeV p+p
collisions as our baseline, which yield good agreement with the ALICE $D_0$
spectrum~\cite{ALICE:2011aa}. Assuming $\delta$-function fragmentation and folding with
EPS09 shadowing~\cite{ALICE:2012ab}, we obtain our initial HQ spectra for the Langevin
simulations. The hydrodynamic medium evolution model has been tuned to available hadron
data at the respective centralities, reproducing fairly well charged-hadron multiplicities,
spectra and $v_2$ at $T_{\rm kin}$, and $\Omega^-$ observables at $T_c$.

\begin{figure}[!t]
\begin{center}
%\vspace{0.2cm}
%\hspace{-0.4cm}
\includegraphics[width=8cm]{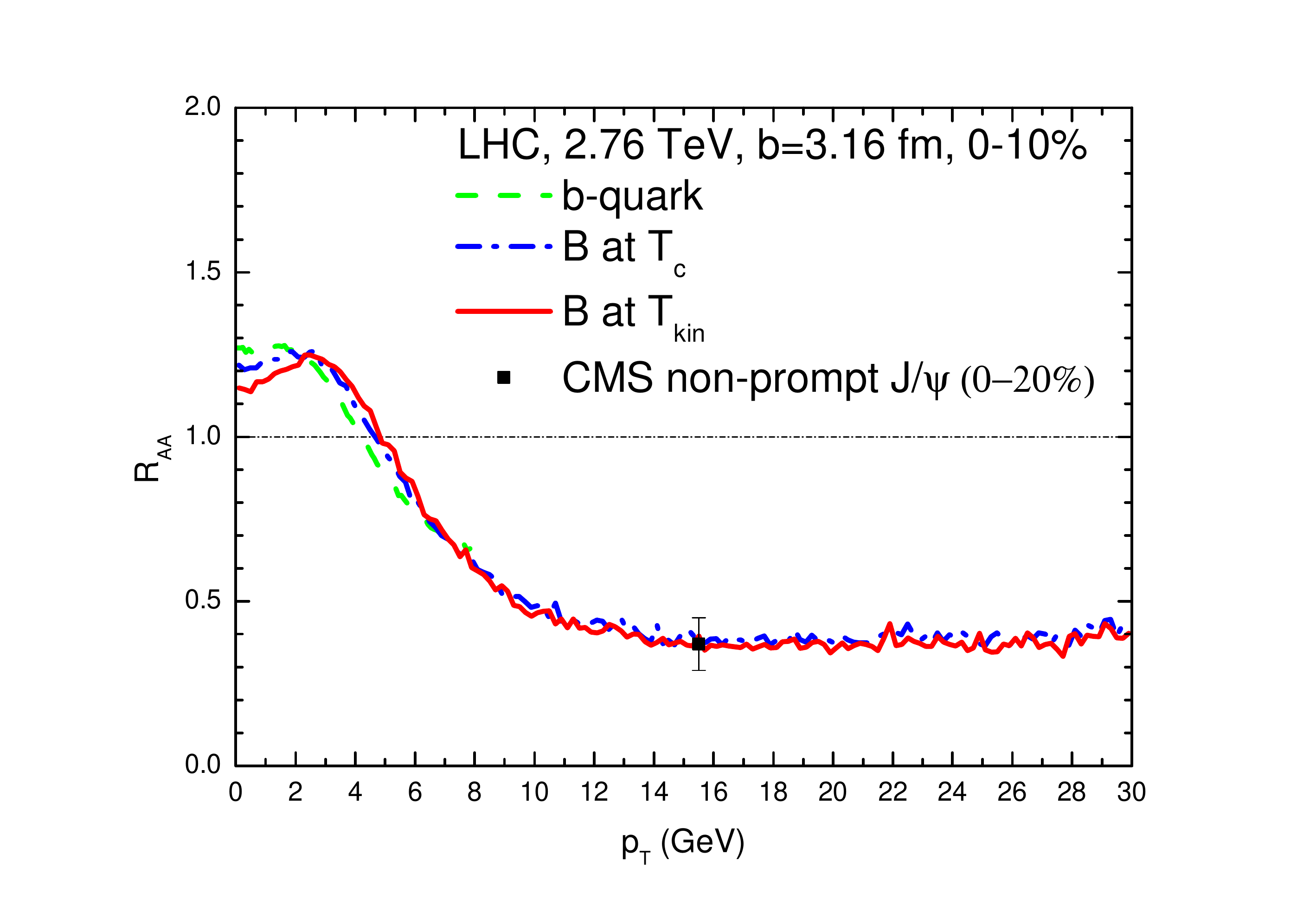}
\includegraphics[width=8cm]{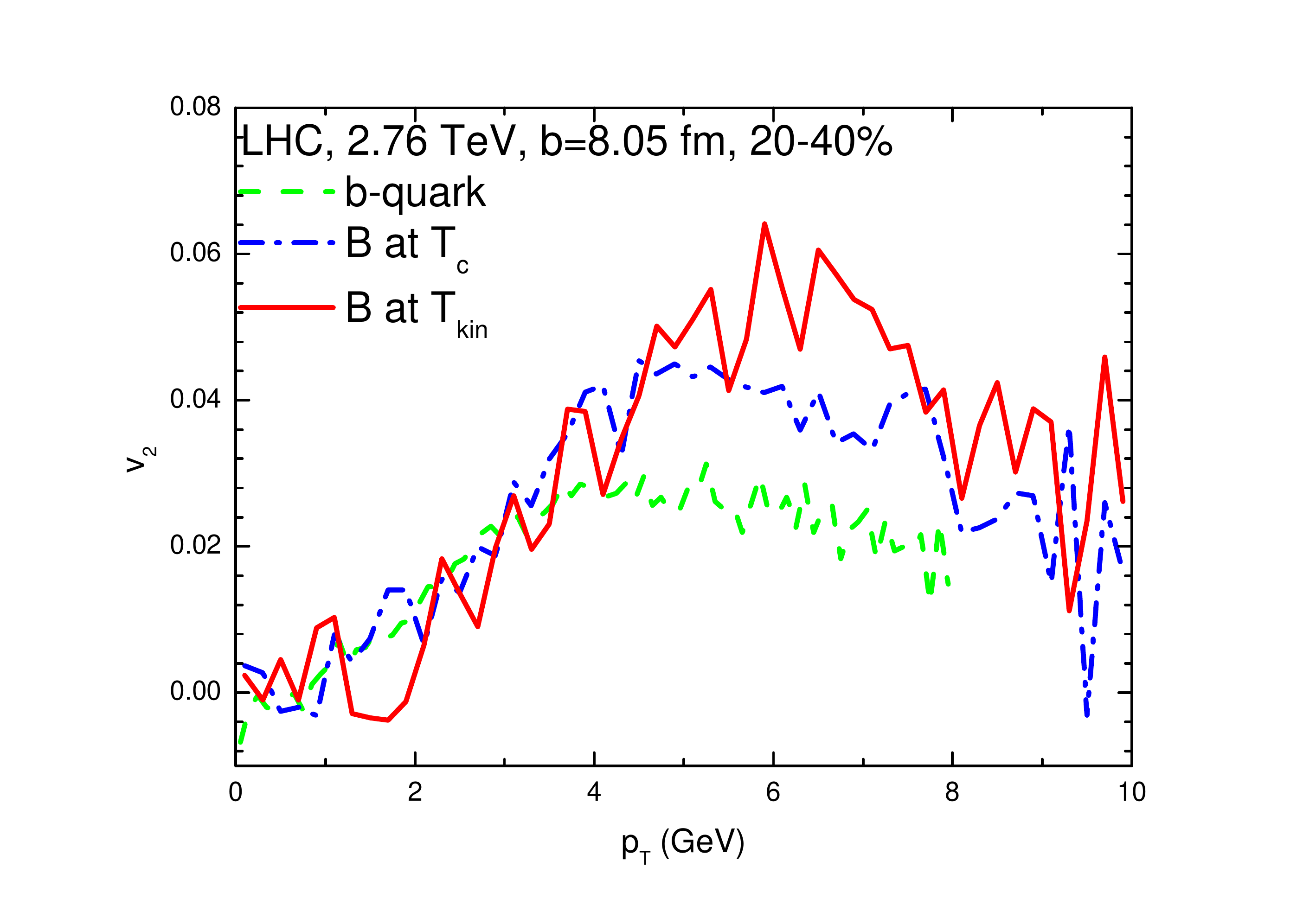}
\end{center}
\vspace{-0.3cm}
\caption{(Color online) $B$-meson $R_{\rm AA}$ (left panel) and $v_2$ (right panel) for
Pb+Pb ($\sqrt{s_{\rm NN}}$=2.76\,TeV) collisions. No shadowing is assumed. The average $p_T$ of the CMS datum~\cite{Chatrchyan:2012np} on non-prompt
$J/\psi$ has been rescaled by $m_B/m_{J/\psi}$.}
\label{LHC_B}
\end{figure}
In Fig.~\ref{LHC_D} we display our calculated $D$-meson $R_{\rm AA}$ and $v_2$ for central
and semi-central collisions, respectively. The flow bump in the $c$-quark $R_{\rm AA}$ at low
$p_T$ is amplified via coalescence at the $D$-meson level. Diffusion of $D$ mesons in the
hadronic phase slightly reduces the $D$-meson $R_{\rm AA}$ at high $p_T$, and increases the
$v_2$ by up to 20\%. Our $D$-meson $R_{\rm AA}$ appears to be systematically slightly too high
but it reproduces the $p_T$-shape of the data~\cite{ALICE:2012ab} rather well, largely induced
by the falling momentum-dependence of the non-perturbative $c$-quark relaxation
rate~\cite{Riek:2010fk,Huggins:2012dj}. The deviation from the data possibly indicates
a missing radiative energy-loss contribution. For the $D$-meson $v_2$ we note that
$c$-quark diffusion through the QGP only accounts for about half of its final value.
Coalescence with light quarks and $D$-meson diffusion in hadronic phase cannot be neglected.
Again, the apparent underprediction of the ALICE data, especially at higher $p_T$ where the
the momentum-dependence of the elastic $c$-quark relaxation rate drops significantly,
might call for an inclusion of radiative contributions. We note that such an assessment is
only possible in the absence of phenomenological $K$-factors.

The $B$-meson results, shown in Fig.~\ref{LHC_B}, show a substantial suppression of
$R_{\rm AA}\simeq0.4$ at $p_T>10$\,GeV, while the $v_2$, reaching up to 5\%, indicates
less collectivity (and thus less thermalization) than for $D$'s. The $B$-meson
modifications also impact $J/\psi$'s via a so-called ``non-prompt" feeddown. In this
context it is interesting to quote the suppression of non-prompt $J/\psi$
of the CMS collaboration~\cite{Chatrchyan:2012np}, which is $\sim$0.37$\pm$0.08 at an average
$p_T^{J/\psi}\simeq 9.3$\,GeV, in good agreement with our $B$-meson result at a correspondingly
somewhat larger parent momentum.

%%%%%%%%%%%%%%%%%%%%%%%%
\section{Conclusions}
%%%%%%%%%%%%%%%%%%%%%%%%
We have evaluated heavy-flavor probes at RHIC and LHC within a  non-perturbative framework
for bulk evolution and microscopic interactions in the diffusion process~\cite{He:2011qa}. Initial
comparisons to $e^\pm$, $D$- and $B$-meson observables at RHIC and LHC are encouraging. \\

%%%%%%%%%%%%%%%%%%%%%%%%%%%%%%%%%%%%%%%%%%%%%%%%%%%%%%%%%%%%%%%%%%%%%%%
{\it Acknowledgments.---}
 %\label{sec_acknow}
%%%%%%%%%%%%%%%%%%%%%%%%%%%%%%%%%%%%%%%%%%%%%%%%%%%%%%%%%%%%%%%%%%%%%%%
This work was supported by the U.S. National Science Foundation
(NSF) through CAREER grant PHY-0847538 and grant PHY-0969394, by the
A.-v.-Humboldt Foundation, and by the JET Collaboration and DOE grant
DE-FG02-10ER41682.

%% References


\bibliographystyle{elsarticle-num}



\end{document}